\documentstyle[12pt,a4]{article}
\renewcommand{\baselinestretch}{1.6}
\begin{document}

\title{Compositional Studies of Ancient Copper from Romanian
Territories}
\author{Agata Olariu$^{\star}$
C. Besliu$^{\circ}$, M. Belc$^{\oplus}$,\\
 I. V. Popescu$^{\star}$, T. Badica$^{\star}$,\\
 Gh. Lazarovici$^{\diamond}$\\
{\em $^{\star}$National Institute for Physics and Nuclear Engineering,}\\
{\em Bucharest, Romania}\\
{\em $^{\circ}$Faculty of Physics, Bucharest}\\
{\em $^{\oplus}$Physics Department, Ovidius University, Consta\c ta, Romania}\\
{\em $^{\diamond}$History Museum of Transylvania, Cluj-Napoca, Romania}}
\date{}

\maketitle
     
\begin{abstract}
Ancient copper objects from Romanian Territories have been analyzed by 
neutron activation analysis. A series of elements is determined: Au, Ag, As,
Co, Cr, Fe, Hg, Ni, Zn, Sb, Sc, Se, Sn. 
Using mathematical dendograms some classifications and correlation  have been
established. 
\end{abstract}
\newpage
\section{Introduction}
The importance of analyses of archaeological objects is well known, 
the elemental composition 
coming to complete the knowledge on the archaeological objects:
style, typology, place of discover, culture, dating, data got by pure
historical arguments.
Processing the compositional scheme of historic objects one can obtain
the identification of the materials and different characterizations
and classifications of the founds.\\
As concerns the ancient copper, the elemental structure shows first of all
the bronzes: copper + arsenic, copper + antimony, copper + tin or incidentally
copper + zinc.  Trace elements could also help to the understanding of copper 
items: objects with closed elemental composition could be put in cultural or 
temporal synchronization.
From the all the analytical methods for copper neutron activation analysis
(NAA)  seems to
be the best single method $^{1}$
%[Pernicka],
 offering enough analytical capabilities
to provide a good data base.
For provenance studies, which it is a very complicated problem the method to
apply is the atomic mass spectroscopy:  what it remained 
constant from the ores
to the final metal objects is the ratio of lead, determined by the atomic mass
spectroscopy: Pb$^{208}$/Pb$^{206}$ and Pb$^{207}$/Pb$^{206}$ $^{2-5}$.\\
In this paper we have analyzed by neutron activation analysis a lot of 41 
Neolithic copper objects: axes, needles, daggers and others objects from the 
National History Museum of Transylvania, Cluj-Napoca, Romania.  
In Table 1 it is given the list of analyzed samples.

\section{Experimental method of analysis}

{\bf Sampling}. First some corroded parts have been
removed from the surface of copper objects, the corroded material having a
totally different elemental composition from that of the body of the object. 
Then 
samples of 10-50 mg have been cut with a hard vidia knife from the object 
body and
after that washed with different solvents: acetone, benzene, ether to avoid the
impurities from the surface of the item and the protective varnish, added 
in the museum.\\   
{\bf Irradiation of medium periods}. Samples have been put in polyethylene 
foils and
irradiated at the rabbit system of the nuclear reactor VVR-S, from NIPNE 
Magurele, Bucharest at the flux of 
$\approx$1.2$x$10$^{12}$neutrons/cm$^{2}\cdot$sec, for 30 minutes.
Copper being in majority it was strongly activated so that the induced
radioactivity in the samples could be measured only after 4-5 days . 
Natural cooper has 2 isotopes:
Cu$^{63}$ and Cu$^{65}$ which by the reaction
(n, $\gamma$) give the radioisotopes 
Cu$^{64}$  T$_{1/2}$=12.74 h and Cu$^{66}$ with T$_{1/2}$=5.10 min. 
After a cooling time of 4-5 days, in the gamma spectra of the samples, the
activity coming from the photopeak of 1345.8 keV (0.0048) of 
Cu$^{64}$ is small enough and permits to remark  other elements, present
in the cooper matrix. The samples have been measured 1000 s at a spectrometric
chain using a Ge(Li) detector of 135 cm$^{3}$ and an analyzer of 4096 channels
coupled at a PC. The system gave a resolution of 2.7 keV at 1.33 keV 
(Co$^{60}$).
We observed the elements: Au, As, Cu and Sb.\\
{\bf Long time irradiation}: The samples of copper  have been wrapped in 
aluminum
foil and put it in a quartz phial together with metallic spectroscopic pure 
standards, 
copper and  nickel and irradiated at the vertical chain of the reactor, 
at a flux of 
$\approx$10$^{13}$neutrons/cm$^{2}$$\cdot$sec, for a period of time of 40 h.
After a cooling time of 2 weeks, we measured the $\gamma$ activity of the
samples at the same spectrometric chain, for 3000 s.
We have determined the following elements: Sb, Ag, Co, 
Cr, Fe, Hg, Ni, Se, Sn.\\
{\bf Cobalt}. Cobalt was determined in the copper object 
using the isotope 
Co$^{60}$ got in the reaction:
     Co$^{59}$(n, $\gamma$)Co$^{60}$. 
Co$^{60}$ is also produced by the reaction: 
 Cu$^{63}$(n, $\alpha$)Co$^{60}$ 
which is important enough in this situation, when the element
copper is the major element ($\approx$99\%).
So that  C$_{Co}$=C$_{total}$ - C$_{Co-Cu}$, where C is the concentration.
Another correction made in the calculus of the cobalt concentration in the
copper samples is that the used cooper standard contains also traces of cobalt.
It was determined that a standard of pure copper has a content of minimum 4 ppm
of cobalt.
Also in the gamma background of the experimental room it were observed the
peaks at the cobalt energies of 1773.2 keV  and 1332.5 keV; therefore from the
respective photopeaks area it was subtracted the area given by the background,
measured for the same period of time as the sample.\\
{\bf Mercury} was determined from the $\gamma$ ray of 279.2 keV and intensity
(81.5\%) of  Hg$^{203}$ with
T$_{1/2}$=46.60d. This ray is interposed with the $\gamma$ ray from
Se$^{75}$, of  279.5 
keV, and intensity 0.25. Therefore the contribution of the mercury must be
extracted from the peak of peak of 279 keV: 
N$_{Hg^{203}}$=N$_{total 279keV}$ - N$_{Se^{75}}$, 
where N$_{Hg^{203}}$ is the counting rate in the peak of 279 keV, 
given by mercury contribution.
N$_{total 279keV}$ is the total counting rate 
in the peak of 279 keV\\
N$_{Se^{75}}$ is the counting rate in the peak of 279 keV, due
to the selenium presence.\\
It was used as reference the selenium peak from the energy 
264.7 keV of  intensity of 0.5658 :\\
$\epsilon_{264keV}\cdot$ s$_{264keVSe^{75}}\cdot$ N$_{264keVSe^{75}}$=
$\epsilon_{279keV}\cdot$ s$_{279keVSe^{75}}\cdot$ N$_{279keVSe^{75}}$
where:
$\epsilon_{264keV}$ is the efficiency of the  detector from 264 keV, 
$\epsilon_{279keV}$, is the efficiency  at the  energy of 
279 keV, 
s$_{264keVSe^{75}}$, the intensity of the  line of 264 keV of Se$^{75}$, 
s$_{279keVSe^{75}}$, the intensity of the line of 279 keV of Se$^{75}$\\
{\bf Nickel}. The concentration of nickel was measured by the isotope 
Co$^{58}$ (T$_{1/2}$=71.3 d). Nickel was a exception by the fact that it was 
determined by the reaction 
Ni$^{58}$(n, p)Co$^{58}$, unlike the other elements determined by the reaction
(n, $\gamma$). 
In the Table 2 are given the results of activation analysis for the 
Prehistoric copper
object, from the National Museum of History from Cluj-Napoca.
The concentrations are given in ppm, and when an element was determined 
in a quantity larger than 10000 ppm, its concentrations was expressed 
in percents, using the notation of \%. The measured errors were the statistical
errors and were in mean of $<$10\%. 
\begin{center}
\section{Results and discussions}
\end{center}
First of all  one distinguishes from the Table 2 the samples of copper 
which contain also other elements than copper, in concentrations higher of 
1\%, the copper based-alloys. 
So that we can distinguish the items:\\
LG10 1.11\% Sn, 
LG14 2.61\% As, 
LG15 1.68\% As, 
LG22 2.82\% Sn + 2.07\% Zn,
LG41 1.53\% As.\\
{\bf Arsenical-copper}
The presence of arsenic in the samples of ancient copper is shown in Fig.
1, the analyzed objects could be structured on 4 levels of concentrations:\\
- concentrations of As $<$ 100 ppm: LG1, LG9, LG11, LG12, LG13, LG24, LG31\\
- concentrations between 100 ppm and 1000 ppm: LG5, LG8, LG21\\
- concentrations between 1000 ppm and 1\%: LG10, LG16, LG22, LG37, LG38, LG39\\
- concentrations $>$ 1\%: LG14, LG15 and LG41\\
The concentrations of arsenic could be interpreted in the sense that 
the samples are made
manufactured by the ore of type reduced, like sulfurs 
and not deliberated alloyed. The specified ores needs a higher degree 
of metallurgy level to produce the metal.
The fact that the items LG14, LG15, LG41 which contain arsenic in 
concentrations higher of 1\% to be manufactured from a reduced ore type is 
possible with a
probability of 3\%$^{6}$, in the rest of chances the  arsenic being added 
intentionally, consciously to obtain  some artifacts of copper with superior
properties than those items of pure copper, the arsenical bronzes$^{7,8}$.\\
As concerning the samples LG16 and LG37, LG38, LG39 with a content of arsenic
of about 1000 ppm, one can do the same discussion but the probability of
bringing in close association  of the object to a ore of reduced type is in
this case  13\%. 
Zvi Goffer$^{9}$ gives another interpretation and method to establish the
affiliation of studied objects at the type of aresnic-copper. 
The Sb and Ag contents of some arsenical copper objects analyzed by Zvi Goffer
are directly proportional to the As content. As, Sb and Ag are associated with
Cu only in sulfoarsenate ores. The presence of
these elements is thus indicative that sulfoarsenate ores must have been
deliberately added during the manufacturing process.
For the  analyzed Neolithic copper from Transylvanian territories we have
represented in Fig. 2 the two dependencies indicated in ref.7:
C$_{As}$ - C$_{Ag}$ and C$_{As}$ - C$_{Sb}$, considering only the objects 
with C$_{As}$$>$1\%. One can observe that the samples LG14
and LG15 respect the proportionality of arsenic-silver, respectively
 arsenic-antimony, and one could make the interpretation that in the process
of the manufacturing of the axes LG14, Dragu and LG15 Hoghiz, it was added a
sulfoarseniate ore deliberately, to obtain axes of higher durity at a smaller
temperature. 
 The dagger LG41 does not respect this rule.

{\bf The axe LG10 and the chisel LG22} have about the same composition, they 
are
tin-copper alloys, which are a superior type of bronze and they have also 
relatively the 
same values of concentrations for the others elements: Au, As
and Sb, Ag and Ni.
Although the axe LG10 contains concentrations of percents of zinc, the value 
C$_{Zn}$=1310 ppm is also a big concentration in comparison with the values of
background for zinc, of the others copper objects, so that one could say that
also by the element Zn
LG10 is synchronized with the object LG22, which contains
C$_{Zn}$=2.07\%. Zinc is reported to be present in the copper object only 
accidentally.
In the literature are given only some metallic founds of zinc$^{7}$: at 
Dordos, also in  Transylvania.
The clusterisation of the 2 objects: axe LG10 and chisel LG22 
on all the diagrams points out their strong association and indicates their
possible common origin and manufacturing place.

As concerns  {\bf the axes from Valcele}, the request was to identify that
the axes LG32 and LG33 belong to the Valcele hoard or are fakes, having  
a very different macroscopic aspect: these items have a dark brown color 
but the other axes  from the same collection are covered by a large layer of 
light blue-greenish copper oxides.  
The elemental analyses show that the samples LG26, LG27, LG28, LG29, LG31, 
LG32, LG33 form o homogeneous group. The axe LG31 only contains a concentration
of 74 ppm. The axes LG32 and LG33 exhibit values of concentrations very closed
of the others axes from Valcele. It could answer in the affirmative that the 
axes
LG32 and LG33 are authentics. \\
The 2 objects found in Ariusd  {\bf the Needle LG40 and the Dagger LG41}
are seemingly very different: the Needle is an unalloyed copper and the Dagger 
is prominent by its content of arsenic 
C$_{As}$=1.53\%, already discussed above, and also by its 
high concentration of Ag: 
C$_{Ag}$=6230 ppm. However the other trace elements: Sb, Se, 
Hg, Ni and especially Fe and Zn suggests the close association of
 these 2 objects.
One could interpret the results of analyses for the Ariusd objects as follows:
the needle and the Dagger seams to be manufactured from the same ore but  
it was added intentionally another ore with arsenic and silver.

The traces of Au and Ag in copper objects were represented in Fig. 3 
%$^{12,13}$,
It is known that there is a proportionality of the silver and gold 
concentrations
both in the "sulfuric" and "oxides" copper ores$^{2,4}$
Our samples follow relatively well the proportionality
C$_{Ag}$ with  C$_{Au}$ with the exception of the dagger from Ariusd.\\
{\bf Pendantivs "Large spiral", "Middle spiral" and "Small spiral"} 
found 
in Cheile Aiudului, the samples LG37, LG38, LG39 present a similar structure: 
they
are arsenic-bronzes and they are well grouped on the elements: Sb, 
Ag and Ni.\\
Sb presents random values of concentrations in the frame of analyzed
samples. It is known that the antimony it is used to replace the tin. 
$^{10,11}$. One can notice for Sb the high values of concentration for
the pendantivs from Cheile Aiudului, the samples 
LG37, LG38, LG39, and also for the objects  LG10 and LG22.\\
{\bf Dendograms}. A study of the correlation of all the analyzed objects
from the point of view of the whole compositional scheme was done using the 
dendogram tools.
$^{12,13}$. In Fig. 4  it is represented the dendogram 
of interdependence between the elements in the cooper samples. It comes in 
to view the levels of
concentrations: I: Au is correlated with Ag the probability P=19.5\%, 
II: Au, Ag, As and Ni, with P=17.0\%, 
III: Au, Ag, As, Ni and Sb with P=14.6\%.
In Fig. 5 it is shown the  dendogram of correlation 
between all the analyzed copper objects from the point of view of all
concentrations. 
These mathematical clustering must be seen cautiously and the obtained
associations must be interpreted only together with historic data: place
of found, dating, culture, typology $^{14}$.

\newpage
{\bf \large References}\\
\noindent
1. E. Pernicka, Nucl. Instr. Meth., {\bf 14} (1986) 24-29\\
2. 2. G. A. Wagner et al., "Arc\"{a}ometllurgische Untersuchungen in
Nordwestanatolien", Vol. 31, 1984\\
3. Thomas C. Seeliger et al., Arc\"{a}ometallurgische Untersucuchungen in Nord -
und Ostanatolien, Jahrbuch der R\"{o}misch-Germanisches Zentralmuseums, Vol.
32, 1985\\
4. A. G. Wagner et al., Geochimische und Isotopische Charakteristika
fr\"{u}her Rohstoffquellen f\"{u}r Kupfer, Blei, Silber und Gold in der
T\"{u}rkei, Jahrbuch der R\"{o}misch-Germanisches Zentralmuseums, Vol. 33, 1986\\
5. E. Pernicka, Erzlagerst\"{a}ten in der \"{A}g\"{a}is und ihre Ausbeutung im
Altertum, Jahrbuch der R\"{o}misch-Germanisches Zentralmuseums, Vol. 34, 1987\\
6. A. M. Friedmann, "Copper artifacts: Correlations with Source Types of
Copper Ores", Science, Vol 152, 1966\\
7. R. F. Tylecotte, "The technique and development of early copper smelting", 
A History of Metallurgy, Ed. The Metals Society, 1978\\
8. J. W. Mellor, A Comprihensive Treatise on Inorganic and Theoretical Chemistry,
Vol. III, The History of Copper, 1956\\
9. Z. Goffer, Archaeolgical Chemistry, Chemical Analysis, Vol. 55, Eds. P.J.
Elving, J.D. Winefordner, I.M. Kolthoff, John Wiley \& Sons, 1980, p.211\\
10. P. T. Craddock, "The Copper alloys of Tibet and their Background", Occasional
Paper, British Museum, No. 15, Eds. W. A. Oddy and W. Zwalf (1981)\\
11. P. T. Craddock, "Deliberate Alloying in the Atlantic Bronze Age", Proceedings
of The Fifth Atlantic Colloquium", Dublin, 1978, p.369\\
12. D. Dumitrescu, "Hierarchical Pattern Classification", Fuzzy Sets Systems 28
(1988) p.145-162\\
13. I. Lup'sa and Gh. Lazarovici, Interdiscipl. Research Com. Seminar , 
Cluj Univ., 19 Dec, 1985\\
14. C. Be'sliu, Gh. Lazarovici, A. Olariu, Acta Musei Porolissensis, 
Vol. XVI (1992) p.97\\
\newpage
\pagestyle{empty}
\renewcommand{\baselinestretch}{1.3}
\setlength{\oddsidemargin}{+1cm}
\scriptsize
\begin{table}[h]
\newlabel{}
\caption{{\bf Table 1}.} \\
% List of anlyzed copper objects from History Museum of Transylvania, by NAA}\\

\begin{tabular}[h]{lcrc}
\hline
Sample & Copper object & Nr. inv. & Place of found\\
\hline
\hline\\
LG1     &   Axe &   P.837 & Ugru'tiu\\
LG2      &  Axe-Pick-axe &    P.838& Mico'slaca\\
LG3     &  Axe-Pick-axe       &    P.839& Cetatea de Balt'a\\
LG4	& Pick-axe     & P.840 & 'Sincai-Pogani\\
LG5&  Battle Pick-axe & P.841& unknown\\
LG6&  Axe & P.842&unknown\\
LG7&  Axe-Pick-axe & P.843 &Lacu, jud. Cluj\\
LG8&  Axe-Pick-axe & P.844 & region Cheile Turzii\\
LG9&  Axe-Pick-axe & P.845 &unknown\\
LG10& Axe & P.846 &unknown\\
LG11& Axe-Pick-axe	& P.847 &unknown\\
LG12& Axe  & P.848 &col. Z. Torma\\
LG13& Pick-axe	& P.850 &unknown\\
LG14& Flat axe     & P.851& Dragu\\
LG15& Chisel     & P.852 & Cold'au, zona Dej\\
LG16& Chisel     & P.853  &Hoghiz\\
LG17& Battle axe      & P.854 &P'adureni(=Be'seneu)\\
LG18& Pick-axe     & P.855 &Col. Z. Torma\\
LG19& Axe  & P.856 &unknown\\
LG20& Axe     & P.857 &unknown\\
LG21& Axe & P.859 &Col. Z. Torma \\
LG22& Flat Chisel      & I 2978      &unknown \\
LG23& Bracelet     & P.?      & V 9558Col. Z. Torma\\
LG24& Bracelet     & P.860       & Cata, C. Petre'sti\\
LG25& Needle  &    & Balomir\\
LG26& Axe  & In 11 & V\^{a}lcele\\
LG27& Axe  & 10470 & V\^{a}lcele\\
LG28& Axe  & 10412 & V\^{a}lcele\\
LG29& Axe  & I 10469& V\^{a}lcele\\
LG30& Bead & 8445   & Decea Mure'sului\\
LG31& Axe  &        & V\^{a}lcele\\
LG32& Axe  &	& V\^{a}lcele\\
LG33& Axe &	& V\^{a}lcele\\
LG34& Small needle   &VI 1399=JUNGH & Decea Mure'sului\\
LG35& Bead    & VI 1293    & Decea Mure'sului\\
LG36& Necklace VI     & 1300=JUNGH 9024           & Decea Mure'sului\\
LG37& Large spiral  &           & Cheile Aiudului\\
LG38& Middle spiral &         & Cheile Aiudului\\
LG39& Small spiral  &           & Cheile Aiudului\\
LG40& Needle         &   III 1789 & Ariu'sd, Covasna\\
LG41& Dagger& III 378 & Ariu'sd, Covasna\\
   &          &         &               \\
\hline     
\end{tabular}
\end{table}

\newpage
\pagestyle{empty}
\renewcommand{\baselinestretch}{1.3}
\scriptsize
\setlength{\oddsidemargin}{0cm}
\begin{table}[h]
\newlabel{}
\caption{\bf Table 2}.\\

\begin{tabular}[h]{lrrrrrrrrrrrrrr}
\hline
Sample& Au & As & Sb & Se & Hg & Cr & Ag & Ni &
Sc & Fe & Zn & Co & Ta & Sn \\
\hline
\hline
LG1&	0.52	&	22	&	8.63	&	7.6	&	2.74	&	$<$2.4	&	32.7	&	99	&	0.039	&	150	&	19.2	&	--	&	0.13	&	180	\\
LG2&	$<$0.54&	--	&	2.12	&	$<$2.4	&	1.28	&	--	&	8.27	&	68	&	--	&	--	&	15.8	&	9.3	&	0.15	&	--	\\
LG3&	0.81	&	--	&	11.4	&	24	&	--	&	--	&	47.4	&	40	&	--	&	--	&	--	&	6.2	&	1.18	&	--	\\
LG4&	0.69	&	--	&	4.8	&	1.5	&	2.35	&	--	&	21.6	&	35.5	&	0.021	&	130	&	14.3	&	--	&	--	&	100	\\
LG5&	1.9	&	110	&	2.94	&	2.9	&	2.49	&	--	&	8.04	&	40.7	&	0.019	&	22	&	12	&	--	&	0.2	&	60	\\
LG6&	0.62	&	--	&	43.5	&	77.4	&	2.39	&	--	&	119	&	48	&	--	&	--	&	39.4	&	--	&	--	&	260	\\
LG7&	8.1	&	--	&	11.9	&	$<$5.1	&	1.9	&	--	&	38.2	&	100	&	--	&	--	&	14	&	--	&	--	&	403	\\
LG8&	0.66	&	127	&	6.72	&	$<$5.1	&	$<$1.7	&	--	&	37.5	&	42	&	0.031	&	500	&	35.4	&	--	&	--	&	390	\\
LG9&	0.39	&	54	&	3.88	&	2	&	0.43	&	--	&	31.6	&	41.4	&	0.087	&	370	&	20.7	&	--	&	--	&	96	\\
LG10&	12.7	&	2820	&	1220	&	24	&	0.8	&	--	&	913	&	370	&	0.198	&	--	&	10	&	410	&	--	&	1.11\%	\\
LG11&	0.69	&	14	&	4.34	&	3.6	&	--	&	--	&	16.4	&	33	&	0.014	&	340	&	18	&	5.2	&	0.29	&	--	\\
LG12&	$<$0.75	&	24	&	1.24	&	2.5	&	--	&	--	&	5.4	&	47.1	&	0.03	&	--	&	13	&	6.1	&	0.18	&	--	\\
LG13&	--	&	48	&	1.76	&	--	&	1.9	&	5.5	&	3.5	&	50	&	0.05	&	380	&	23	&	1.6	&	0.4	&	--	\\
LG14&	4.4	&	2.61\%	&	85.5	&	37.4	&	--	&	--	&	163	&	85	&	0.04	&	540	&	28	&	--	&	--	&	--	\\
LG15&	3.4	&	1.68\%	&	20.5	&	48.8	&	--	&	--	&	54.9	&	68.9	&	0.02	&	--	&	19.3	&	--	&	--	&	--	\\
LG16&	9.2	&	6270	&	10.4	&	8.4	&	1.9	&	--	&	378	&	23	&	--	&	--	&	28	&	--	&	0.87	&	--	\\
LG17&	--	&	--	&	2.36	&	11.7	&	--	&	--	&	10	&	53	&	--	&	--	&	29	&	8.6	&	0.97	&	--	\\
LG18&	--	&	--	&	0.72	&	--	&	--	&	--	&	4.8	&	49	&	--	&	400	&	25	&	16.7	&	0.25	&	--	\\
LG19&	$<$1.5	&	--	&	10.5	&	17	&	--	&	--	&	39.3	&	120	&	0.05	&	--	&	36	&	32.2	&	2.31	&	--	\\
LG20&	27.8	&	--	&	5.18	&	89.5	&	1.96	&	--	&	94.6	&	184	&	0.04	&	--	&	21	&	--	&	1.37	&	--	\\
LG21&	11.7	&	550	&	18.3	&	13.9	&	1.28	&	--	&	256	&	--	&	--	&	--	&	34	&	8.8	&	0.53	&	--	\\
LG22&	10.9	&	1540	&	2040	&	--	&	--	&	--	&	925	&	1210	&	--	&	--	&	2.07\%	&	7.8	&	--	&	2.82\%	\\
LG23&	8.8	&	--	&	3.36	&	19.6	&	1.55	&	--	&	381	&	47	&	--	&	--	&	22	&	27.6	&	2.2	&	--	\\
LG24&	0.32	&	19	&	3.1	&	1.4	&	0.43	&	$<$1.6	&	17.9	&	43.9	&	--	&	--	&	31.9	&	--	&	0.19	&	--	\\
LG25&	--	&	--	&	8.53	&	--	&	$<$4.7	&	11	&	39.8	&	110	&	--	&	660	&	30	&	--	&	1.9	&	--	\\
LG26&	1.4	&	--	&	1.23	&	14.4	&	--	&	--	&	69.5	&	47.7	&	0.02	&	--	&	93	&	--	&	0.22	&	--	\\
LG27&	0.8	&	--	&	3.55	&	$<$4.6	&	0.67	&	4	&	31.7	&	56.5	&	0.1	&	--	&	13	&	--	&	--	&	--	\\
LG28&	1.4	&	--	&	4.37	&	13.7	&	0.417	&	--	&	81.5	&	63.9	&	0.03	&	--	&	18	&	11.3	&	0.82	&	--	\\
LG29&	3.2	&	--	&	4.76	&	22.7	&	1.57	&	--	&	114	&	64	&	--	&	--	&	--	&	--	&	--	&	--	\\
LG30&	1	&	--	&	6.69	&	3.8	&	1.7	&	--	&	53.8	&	55.2	&	0.02	&	--	&	26	&	--	&	0.3	&	--	\\
LG31&	3.4	&	74	&	5.88	&	11	&	2.4	&	--	&	60.2	&	51	&	0.03	&	--	&	25	&	--	&	--	&	--	\\
LG32&	1.62	&	--	&	11.8	&	9.3	&	1.65	&	--	&	97.5	&	47.9	&	0.02	&	--	&	19	&	--	&	--	&	--	\\
LG33&	0.9	&	--	&	2.87	&	9.1	&	0.96	&	--	&	63.5	&	24	&	--	&	--	&	15	&	--	&	--	&	--	\\
LG34&	$<$0.4	&	--	&	1.68	&	9.73	&	2.09	&	--	&	6.21	&	32.4	&	0.054	&	150	&	25.3	&	--	&	--	&	--	\\
LG35&	2.76	&	--	&	14.2	&	2.3	&	0.926	&	1.3	&	27.1	&	67.7	&	0.115	&	1140	&	16.8	&	--	&	0.08	&	60	\\
LG36&	--	&	--	&	3.63	&	--	&	8.8	&	--	&	23	&	--	&	0.15	&	1300	&	160	&	--	&	2	&	--	\\
LG37&	25.9	&	3600	&	548	&	181	&	8.46	&	--	&	307	&	330	&	--	&	--	&	--	&	10	&	--	&	--	\\
LG38&	7.5	&	3250	&	307	&	71.5	&	2.5	&	--	&	97.5	&	419	&	0.56	&	--	&	140	&	2	&	--	&	--	\\
LG39&	4.7	&	1300	&	634	&	--	&	55	&	--	&	155	&	150	&	0.08	&	--	&	755	&	12	&	--	&	--	\\
LG40&	--	&	--	&	2.58	&	17.9	&	8.5	&	--	&	4.82	&	173	&	8.03	&	690	&	39	&	2.7	&	--	&	--	\\
LG41&	$<$0.9	&	1.53\%	&	12.3	&	12	&	4.4	&	--	&	6230	&	92	&	--	&	850	&	68	&	--	&	--	&	--	\\
   &            &               &               &               &       &&&&&&&&&\\
\hline

\end{tabular}
\end{table}

\newpage
\large
\pagestyle{empty}
FIGURE CAPTIONS\\

Fig. 1 Diagram of As content in the copper objects, by NAA\\

Fig. 2 Concentrations of As versus the concentrations of Sb or Ag for 
different ancient copper items, by NAA\\

Fig. 3 Concentration of Ag versus the concentration of Au for different 
ancient copper objects, by NAA\\

Fig. 4 Dendogram of interdependence between elements in ancient copper objects\\

Fig. 5 Dendogram of correlation of the copper objects\\

\newpage

TABLE CAPTIONS\\

Table 1 List of anlyzed copper objects from 
History Museum of Transylvania, Cluj-Napoca, by NAA\\

Table 2 Concentration of elements in different Neolithic copper objects, bye NAA.
 The concentrations are expressed in ppm and for the values above 10 000 ppm,
the results are given in \%. The errors are  generally  $<$7\%.

\end{document}